\documentclass{article}
\usepackage{arxiv}

\usepackage[utf8]{inputenc} 
\usepackage[T1]{fontenc}    
\usepackage{hyperref}       
\usepackage{url}            
\usepackage{booktabs}       
\usepackage{amsfonts}       
\usepackage{nicefrac}       
\usepackage{microtype}      
\usepackage{amsmath}
\usepackage{lipsum}
\usepackage{graphicx}
\graphicspath{ {./images/} }

\usepackage{overpic}
\usepackage{epstopdf}
\usepackage{multirow}
\usepackage[table]{xcolor}

\usepackage{siunitx}
\usepackage{hyperref}
\usepackage{subcaption}

\title{Deep learning-based identification of sub-nuclear structures in FIB-SEM images}

\author{
 Niraj Gupta$^*$ \\
  Department of Computer Science\\
  University of Colorado, Boulder\\
  Boulder, CO 80309, USA  \\
  \texttt{niraj.gupta@colorado.edu} \\
   \And
 Eric J. Roberts$^*$ \\
  Molecular Biophysics \& Integrated Bioimaging Division\\
  Center for Advanced Mathematics for Energy Research Applications\\
  Lawrence Berkeley National Laboratory\\
  Berkeley, CA 94720, USA \\
  \texttt{EJRoberts@lbl.gov} \\
\And
   Song Pang \\
   Janelia Research Campus\\ 
   Howard Hughes Medical Institute \\ 
   Ashburn, VA 20147, USA\\
  \texttt{pangs@janelia.hhmi.org} \\
 \And
 C. Shan Xu\\
  Janelia Research Campus\\ 
   Howard Hughes Medical Institute \\ 
   Ashburn, VA 20147, USA\\
  \texttt{xuc@janelia.hhmi.org} \\
   \And
 Harald F. Hess\\
   Janelia Research Campus\\ 
   Howard Hughes Medical Institute \\ 
   Ashburn, VA 20147, USA\\
  \texttt{hessh@janelia.hhmi.org} \\
\And
    Fan Wu\\
    Department of Molecular and Cellular Biology\\
    University of California, Berkeley\\
    Berkeley, CA 94720, USA
\And
    Abby Dernburg\\
    Department of Molecular and Cellular Biology\\
     University of California, Berkeley\\
     Berkeley, CA 94720, USA
\And
    Danielle Jorgens\\
    Electron Microscopy Labs\\
    University of California, Berkeley\\
    Berkeley, CA 94720, USA
\And
 Petrus H. Zwart$^\dagger$ \\
    Molecular Biophysics \& Integrated Bioimaging Division\\
  Center for Advanced Mathematics for Energy Research Applications\\
  Lawrence Berkeley National Laboratory\\
  Berkeley, CA 94720, USA \\
  \texttt{PHZwart@lbl.gov} \\
\And
   Vignesh Kasinath$^\dagger$ \\
  Assistant Professor\\
  Department of Biochemistry \\ 
  University of Colorado, Boulder\\
  Boulder, CO 80309, USA  \\
  \texttt{vignesh@colorado.edu} \\
}

\begin{document}
\maketitle

\def\thefootnote{$*$}\footnotetext{N. Gupta and E. Roberts contributed equally to this work}

\def\thefootnote{$\dagger$}\footnotetext{Corresponding Authors}

\begin{abstract}
    Three-dimensional volumetric imaging of cells allows for \emph{in situ} visualization, thus preserving contextual insights into cellular processes. Despite recent advances in machine learning methods, morphological analysis of sub-nuclear structures have proven challenging due to both the shallow contrast profile and the technical limitation in feature detection. Here, we present a convolutional neural network, supervised deep learning-based approach which can identify sub-nuclear structures with 90\% accuracy. We develop and apply this model to {\it C. elegans} gonads imaged using focused ion beam milling combined with scanning electron microscopy resulting in the accurate identification and segmentation of all sub-nuclear structures including entire chromosomes. We discuss in depth the architecture, parameterization, and optimization of the deep learning model, as well as provide evaluation metrics to assess the quality of the network prediction. Lastly, we highlight specific aspects of the model that can be optimized for its broad application to other volumetric imaging data as well as \emph{in situ} cryo-electron tomography.
\end{abstract}


\section{Introduction}

Focused ion beam scanning electron microscopy (FIB-SEM) is an automated imaging technique capable of probing structure on the micrometer to nanometer scale. FIB-SEM consists of dual beam instrumentation in which a scanning electron microscope works in combination with high-energy ion beams that mill away,  \emph{in situ}, a resin-encased specimen. In this setup, the microscope directly observes the newly revealed ultrathin layer of the sample as it is milled away. The serial images are then collected and aligned to produce a 3d volume image stack. This led to unprecedented $z$-axis resolution in the range of tens of nanometer, resulting in isotropic (i.e. $z$-axis sectional thickness is equal to $x-$ and $y-$axis pixel size), full image resolutions and user-specified site-specific selections of 4\si{\nano\meter}$^3$-sized voxels, achieved in a pioneering study of aldehyde-fixed mouse brain tissue images \cite{wu2017contacts, xu2021open, heinrich2021whole}. This represents an order of magnitude improvement in $z$-axis resolution over the best current SEM-only techniques, which alternate between imaging and cutting away the surface with diamond-tipped knives \cite{hayworth2006automating, wanner2015challenges}. While these and similar cutting strategies were significant advancements for biological imaging – this represented the first instance of a fully automated, \emph{in situ} volumetric EM approach performed by cutting material layers off  \cite{denk2004serial}. Serial block-face methods lose consistency when cutting less than 25 \si{\nano \meter} deep \cite{nalla2005ultrastructural, giannuzzi1999review}, thus limiting both resolution and the acquisition volume. In replacing the diamond knife with the ion beam, FIB-SEM approaches, but does not reach, the state-of-the-art resolution for nanometer-scale 3D imaging of $\sim$1\si{\nano\meter}-sized voxels, achieved by transmission electron microscopy (TEM) where electrons are transmitted through the medium or sample. State-of-the-art cryo-TEM imaging is typically limited to sample thickness of less than  $\sim$ 300 \si{\nano\meter}. FIB-SEM, however, allows for far greater imaging depth and unparalleled levels of automation that is constrained only by the availability of material, time, and operational costs. 

Before being applied in biological settings, FIB-SEM was first proposed and implemented in soft matter physics and has long proven invaluable to material scientists, leading to effective analysis and classification of micro structures in porous media \cite{holzer2004three, kelly2016assessing, wu2019analyses, gostick2019porespy}, characterization of nanopores in coal and other reservoir rocks \cite{fang2019methodology, garum2020micro}, the reconstruction of polymer films \cite{vcalkovsky2021comparison, roding2021three}, and better optimization and design of controlled drug release coatings \cite{fager2020optimization}. FIB-SEM dual beam setups have since been explored for biological imaging, beginning with a FIB used as a cutting device for exposing tissue and gland cells \emph{in situ} to subsequent SEM imaging \cite{drobne2004focused}. Though imaging resolution was limited to relatively shallow depths on the scale of dozens of micrometers, this was a major step forward for FIB-SEM technology and marked the beginning of modern volumetric microscopy \cite{kizilyaprak2019volume}. More recently, research developments in FIB-SEM methodology have matured the technique into an essential tool in the study of cellular biology \cite{hoffman2020correlative, muller20213d, weigel2021er, kizilyaprak2019volume, drobne2004focused, drobne2005electron, vidavsky2016cryo}. These developmental breakthroughs include correcting for anisotropic data slicing with optical flow interpolation \cite{gonzalez2022optical}, accelerating image acquisition times via adaptive sampling that leverages multiple low-dose, quicker scans \cite{dahmen2016feature}, software improvements incorporating back scattered electron detectors with positive sampling bias \cite{xu2017enhanced}, and considerable improvements in hardware and reliability of continuous, long-term operational sessions, upwards of several months \cite{xu2017enhanced, hayworth2015ultrastructurally}. 

In total, the improvements in FIB-SEM imaging have led to a vast increase in the amount of high quality cellular data to analyze. This is perhaps best exemplified in the recent open-source availability on the web repository ``OpenOrganelle \cite{heinrich2021whole}’’ of numerous whole cell atlas and tissue 3D images of 4 \si{nano\meter} isotropic voxel resolution, with some volumes measuring greater than $100,000$ \si{\micro\meter}$^3$ in size \cite{xu2021open}. While FIB-SEM data requires little in terms of post-processing and alignment, the sheer amount of data produced is a major bottleneck as the ever-increasing need for laborious, manual curation of expert annotations and analyses is only exacerbated. To overcome this barrier, there has been a major push towards automation in image analysis pipelines which will be crucial in furthering the understanding of cellular structures and subcellular components \cite{perez2014workflow}. The rapid advances in machine learning approaches can not only accelerate microscopy data analysis but also offer researchers tools that remain significantly more tolerant to noise than traditional computer vision techniques \cite{andrew2018quantified}. Recent applications of machine learning in electron tomography have helped identify ribosomes, proteomes, mitochondria, Golgi, and other contrast rich objects \cite{zeng2019aitom, bauerlein2021towards, li2019automatic, gubins2020shrec, moebel2021deep, de2022convolutional}. Chromatin structures in the nucleus, however, have not been extensively investigated using both traditional user-dependent approaches as well as through the use of machine learning.

In this study, we apply machine learning tools for the automated delineation and pixel-by-pixel segmentation of intracellular structures in FIB-SEM-acquired image volumes of Caenorhabditis elegans ({\it C.elegans}) reproductive eggs. The data is considerably large and detailed, with a FIB-SEM state-of-the-art resolution of $4\times4\times4$ \si{\nano\meter}$^3$ for volumes measuring upwards of $11,250$ \si{\micro/meter}$^3$. This resolution and volume allows us to identify and segment \emph{all} sub-nuclear cellular structures, including the nucleolus, chromosomes, chromosome-encased synaptonemal complexes, and nuclear membrane. Machine learning network classifier training is performed on a small subset of manually curated labels consisting of only 0.1\% of \emph{all} pixels in the full dataset. To the authors' best knowledge, this full nuclear segmentation has not been performed at this resolution in an automated fashion.

The remainder of this manuscript is organized as follows: Sect.~\ref{sect:data} describes the {\it C. elegans} sample preparation and imaging process; Sect.~\ref{sect:methods} details the data pre-processing, neural networks, training parameters, and evaluation metrics used for analysis; Sect.~\ref{sect:workflow} walks the reader through the network training and image segmentation workflow; Sect.~\ref{sect:results} presents machine learning-based segmentation results; and Sect.~\ref{sect:discussion} offers a discussion on future works and the scope of the manuscript. 



\section{The Data} \label{sect:data}

The data used in this study are volumetric images of {\it C. elegans} gonads at three different stages of meiosis I pachytene: early, mid, and late. {\it C. elegans} is generally seen as an exemplary and prototypical organism in the investigation of developmental biology \cite{brenner1974genetics, hodgkin1977mutations}. {\it C. elegans} was in fact the first multicellular organism to have its entire genome sequenced \cite{coulson1986toward, c1998genome}. Studies of {\it C. elegans} have helped behavioral scientists map the neural circuitry that controls touch-induced locomotion \cite{chalfie1985neural}, deduce the functions of certain touch circuitry neurons \cite{chalfie1985neural}, investigate clues related to the evolutionary development of the circadian clock \cite{banerjee2005developmental}, and discover nested neurological dynamics/activity patterns that govern a behavioral hierarchy of motor actions across multiple time scales \cite{kaplan2020nested}. {\it C. elegans} remains the only organism to have the \emph{entirety} of its nervous system, also known as the connectome, mapped out \cite{white1986structure, towlson2013rich, cook2019whole}

\subsection{FIBSEM Sample Preparation}
One Durcupan-embedded {\it C. elegans} gonad sample was first mounted to the top of a 1 mm copper post which was in contact with the metal-stained sample for belter charge dissipation, as previously described in \cite{xu2017enhanced}. The vertical sample post was first trimmed to a small block of 95 x 80 x 150 \si{\micro\meter}$^3$ containing two Regions of Interest (ROI 1-2) from top to bottom. The sample block has a width perpendicular to the ion beam, and a depth in the direction of the ion beam. After FIB-SEM imaging ROI1 and ROI2, the sample was then trimmed to the second block of 80 x 80 x 100 µm3 containing ROI3. The trimming was guided by X-ray tomography data obtained by a Zeiss Versa XRM-510 and optical inspection under a microtome. Thin layers of conductive material of 10-nm gold followed by 100-nm carbon were coated on the trimmed samples using a Gatan 681 High-Resolution Ion Beam Coater. The coating parameters were 6 keV, 200 nA on both argon gas plasma sources, 10 rpm sample rotation with 45-degree tilt.

\subsection{FIB-SEM 3D large volume imaging}

One FIB-SEM prepared {\it C. elegans} gonad sample was imaged sequentially by a customized Zeiss FIB-SEM system (Germini 500) previously described in \cite{xu2017enhanced}, \cite{xu2020enhanced}, and \cite{xu2021open}. The block face of each ROI was imaged by a 250 pA electron beam with 0.9 keV landing energy at 200 kHz scanning rate. The $x$-$y$ pixel resolution was set at 4 nm. A subsequently applied focused Ga+ beam of 15 nA at 30 keV strafed across the top surface and ablated away 4 nm of the surface. The newly exposed surface was then imaged again. The ablation-imaging cycle continued about once every 75 seconds for 9 and 6 days to complete FIB-SEM imaging ROI1 and ROI2, respectively. Such cycle extended to once every 135 seconds for 14 days to image ROI3. The acquired image stack formed a raw imaged volume, followed by post-processing of image registration and alignment via local feature matching using a Scale Invariant Feature Transform (SIFT)-based algorithm \cite{lowe1999object, lowe2004distinctive, yang2018high}. The aligned stack consists of final isotropic volumes of $10 \times 20 \times 30$ \si{\micro\meter}$^3$, $10 \times 20 \times 20$ \si{\micro\meter}$^3$, and $25 \times 15 \times 30$ \si{\micro\meter}$^3$ for ROI1, ROI2, and ROI3, respectively. The voxel size of $4 \times 4 \times 4$ \si{\nano\meter}$^3$ was maintained for each sample throughout entire volumes, which can be viewed in any arbitrary orientations. 
 

\begin{figure}[!htb]
\minipage{0.33\textwidth}
    \centering
    \includegraphics[width=5cm,height=10cm]{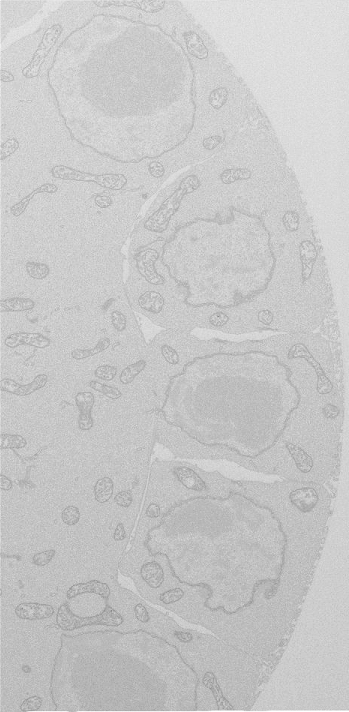}
  
\endminipage\hfill
\minipage{0.33\textwidth}
    \centering
    \includegraphics[width=5cm,height=10cm]{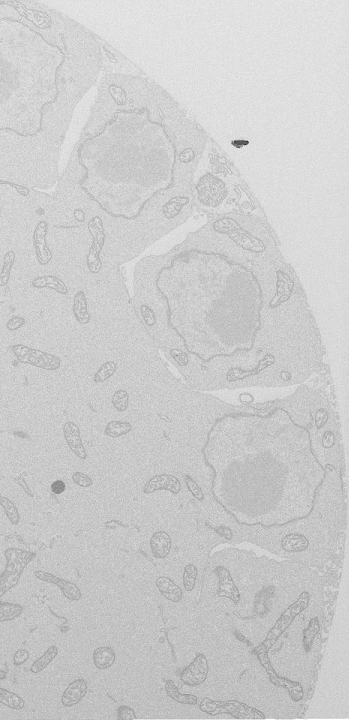}
  
\endminipage\hfill
\minipage{0.33\textwidth}%
    \centering
    \includegraphics[width=5cm,height=10cm]{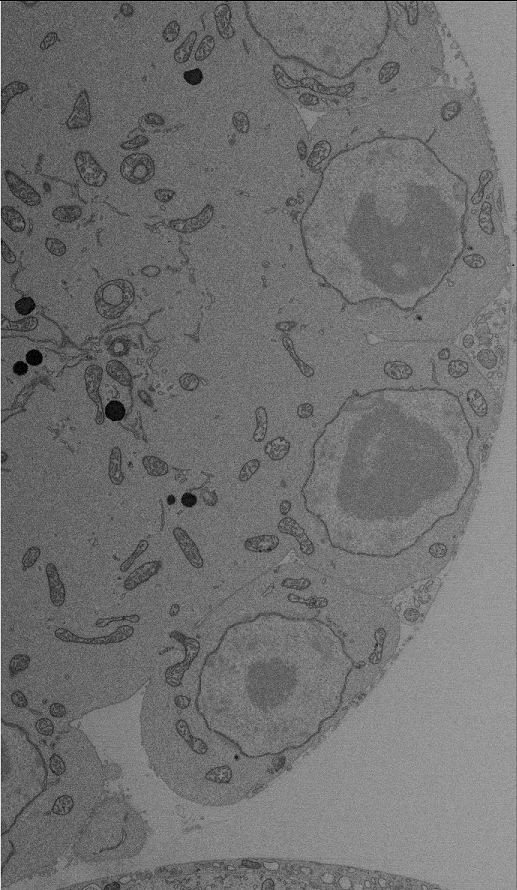}
 
\endminipage
 \caption{Individual FIB-SEM cross sectional slices of \emph{Caenorhabditis elegans}. Pictured are (a) ROI1, (b)ROI2, and (c)ROI3 regions of interest.}
 \label{fig:fib-sem}

\end{figure}




\section{Methods}\label{sect:methods}

\subsection{Data Preprocessing} \label{sect:preprocessing}

The images generated for the FIBSEM images are in TIFF format. ROI1 has 6801 color-scale images that are $5000\times2500\times3$, ROI2 has 5000 images of the same size/dimensionality, and ROI3 has 8837 images that are $3750\times6250\times3$. The nuclei are extracted sequentially from each tiff file using traditional computer vision techniques. Since the nuclear membrane is not continuous due to the presence of nuclear pore complexes which show different contrast compared to the membrane, contour detection cannot be applied directly to the raw images to isolate the nuclei from the 3D volume stack. Instead, we perform image pre-processing by first minimizing image noise using Gaussian blur with an $11\times11$ kernel. Gaussian blur enhances image structure by smoothing out pixel intensities \cite{singhal2017study}, as pixel intensity and brightness are not uniform across the entirety of an image. Second, we perform Gaussian adaptive thresholding to help alleviate remaining inhomogeneous pixel intensities. Thresholding helps extract the margins of nuclei by producing a binary image of the nuclei outline. To fill the gaps in the nuclei boundaries, we first create a $7\times7$ elliptical kernel, which is then used to calculate the morphological gradient \cite{serra1982image, rivest1993morphological}, which highlights stark contrasts in neighboring pixel intensities to form object outlines using the difference between the dilation and erosion morphological operators \cite{lee1987morphologic, na2019filter}. The complete outlines will be fully recognized once the broken edge gaps have been filled. To accomplish this, the approximation method is set to chain approximation, and the retrieval method is set to RETR \textunderscore TREE. The chain approximation method used here stores all the boundary points of the contour \cite{etemadi1992robust, akinlar2011edlines}. The tree retrieval method retrieves all of the contours by traversing over the `chains' of boundary pixels, then reconstructs a full hierarchy of nested contours. These are then filtered to recover elliptical outlines within a specific size range. After filtering, the centroid is used to cluster the nuclei at the same location for any observed contour. Throughout the process of clustering nuclei based on position, a mapping of nuclei position and ID is retained. All of the nuclei collected are stored as JPEG files with a constant size of $1700\times1700$ pixels.

\begin{figure}[!htb] 
\centering
\minipage{0.5\textwidth} 
    \centering
    \includegraphics[width=7cm,height=7cm]{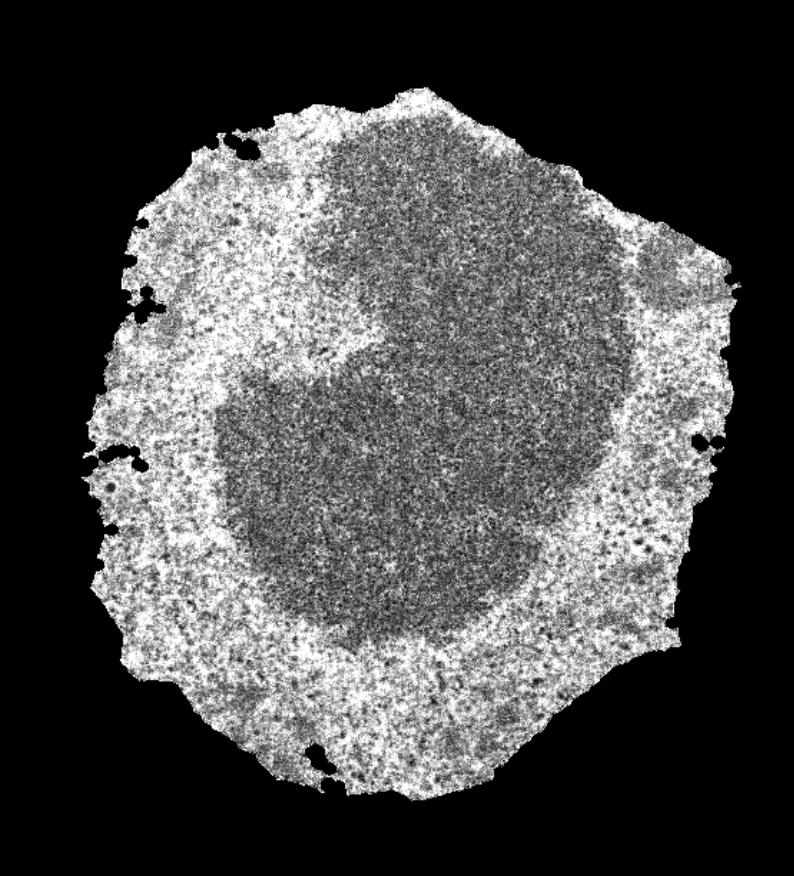}
  
\endminipage
\minipage{0.5\textwidth}
    \centering
  \includegraphics[width=7cm,height=7cm]{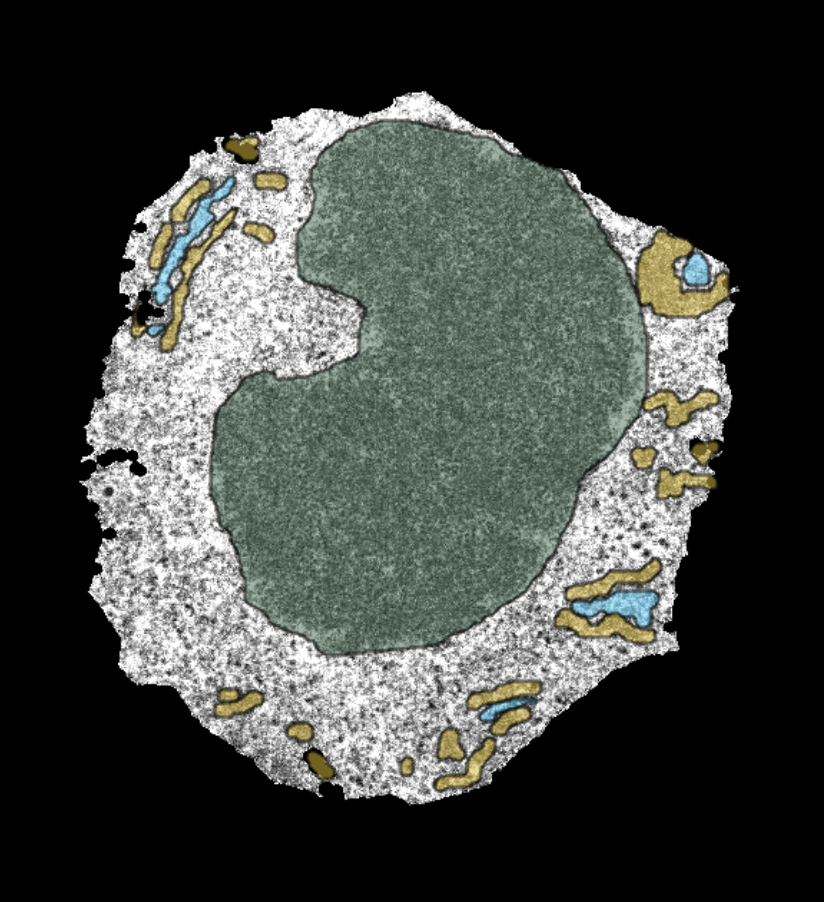}
  
\endminipage
 \caption{(a) Image of an extracted nuclei from ROI1 and  (b) corresponding annotations. Here, green represents nucleolus, blue represents synaptonemal complex, and yellow  represents the chromosome annotations. }\label{fig:annotations}
 

\end{figure}

Figure~\ref{fig:annotations} shows the manual annotations that are created for one of the ROI1 images. The annotations are created using apeer web application (CITE apeer). All the nucleolus are labeled as 1 (green), chromosome as 2 (yellow), and the synaptonemal complex as 3 (blue) with the background as 0 for every image. For the training data, 260 images are randomly sampled and annotated from the generated images of nuclei of ROI1. The generated annotation file are shuffled and processed to generate a new label for rest of the nuclei to improve the performance of the network. 10\% of the data is used as a testing set for cross validation purposes.

\subsection{Convolutional Neural Networks}

Convolutional neural networks (CNNs) are feed-forward, deep learning architectures made up of several connected convolutional layers \cite{fukushima1982neocognitron, lecun1998gradient, simard2003best}. CNNs approximate some underlying unknown function that maps input data to some target domain. In this case of this study, the mapping to be approximated is that of raw input image data to the classification of each pixel to a particular label (chromosome, nucleolus, etc.); i.e. supervised semantic segmentation. As information passes through the network, each convolutional layer convolves the preceding layer's output with an increasing number of two-dimensional convolutional filters, resulting in an intermediate feature map that is passed along as input to the next layer or operation. The additional operations used between adjacent convolutional layers typically consist of nonlinear activation functions and normalization layers which help expedite the learning process, and max pooling to introduce translation invariance \cite{scherer2010evaluation} and reduce computational costs via spatial coordinate downsampling.  

Imperative to the CNN learning process are the convolutional filters. Each filter, typically of size $3\times3$ or $5\times5$, consists of weights to be learned during network training and acts as a smaller receptive field of view that houses some learned feature from the overall image set which may then be re-used and applied to more-complex image reconstruction tasks in the later network layers. This allows for a more-global learning paradigm and deeper, more-parameter efficient neural network architectures than that of more-traditional fully-connected neural networks (FCNNs) \cite{rosenblatt1958perceptron, goodfellow2016deep, xu2019overfitting} in which each individual pixel and connected node is assigned a learnable weight, and individual learned features remain localized to the single spatial coordinates in which they were found, not to be reused anywhere else in the network. 

\paragraph{U-Net}
\label{sect:u-net}

\begin{figure}[h]
\centering
    \includegraphics[width=.95\textwidth]{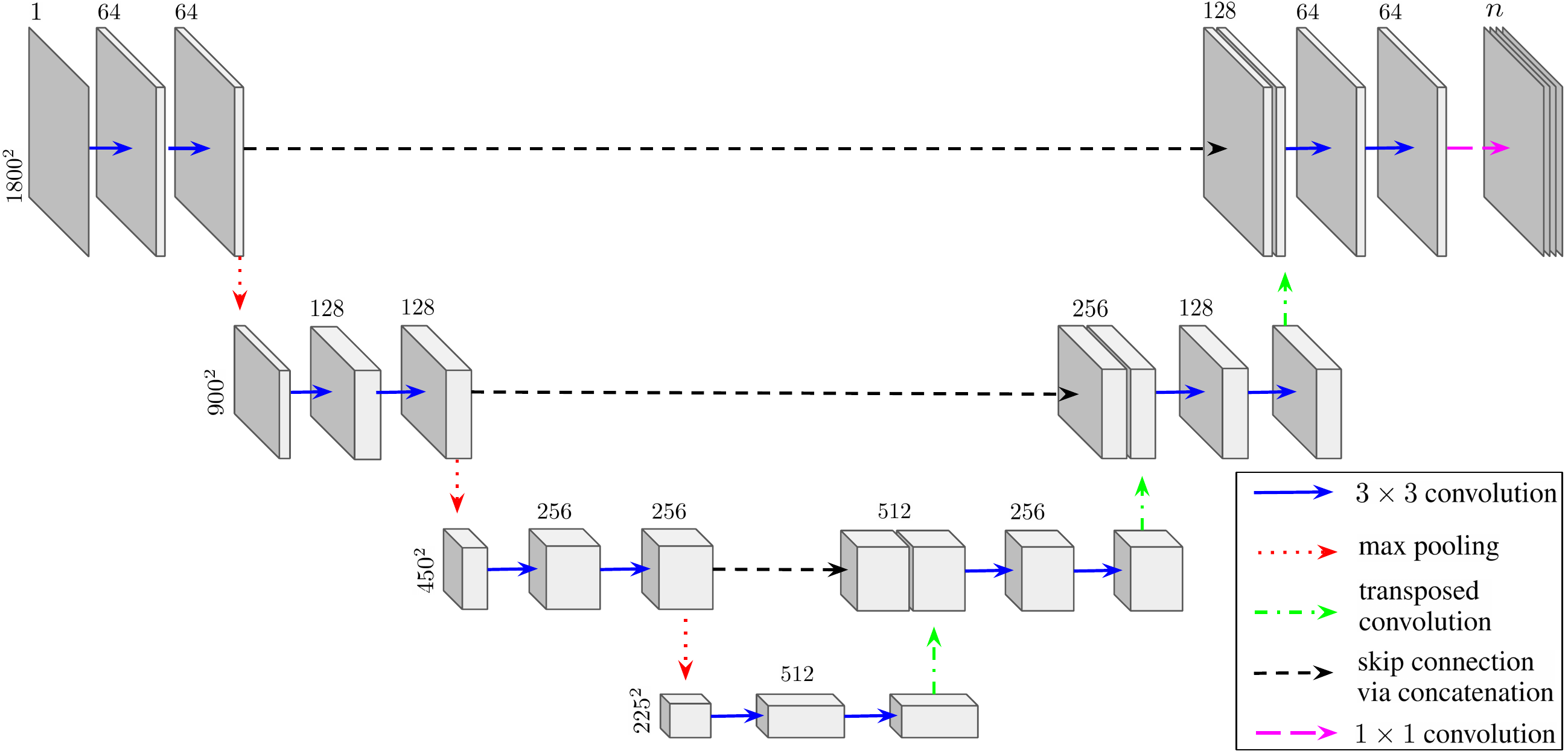}
    \caption{Schematic of a four-layer deep U-Net showing individual operations and intermediate feature map dimensions. In the left encoder-half, max pooling operations (red dotted arrows) halve spatial dimensions and convolutional operations (blue arrows) double the number of channels and filters at each subsequent layer. In the right decoder-half, convolutions decrease the number of channels and transposed convolutions (green dashed-dotted arrows) upsample the spatial dimensions. Skip connections (horizontal dashed arrows) join the two network halves. Lastly a single convolution with filter size $1 \times 1$ (purple long dashed arrows) reduces the network output to any desired $n$-number of channels, which in this case is four (one for background and three for the nucleolus, chromosomes, and synaptonemal complex). }
    \label{fig:unet_schematic}
\end{figure}

The main neural network model we implement for nucleolus segmentation is the U-Net \cite{u_net}, a deep convolutional network first used for pixel-by-pixel segmentation of biomedical images \cite{u_net}. Inspired by convolutional autoencoders \cite{mcclelland1987parallel, demers1992non}, the U-Net, pictured in Fig.~\ref{fig:unet_schematic}, is a symmetric encoder-decoder system made up of two distinct halves: the beginning contractive encoder-half on the left of Fig.~\ref{fig:unet_schematic} aims to capture contextual information and detect important image features with an increase in the number of convolutional channels and corresponding filters, while the expansive decoder-half on the right projects the learned features back into the higher resolution image space to reconstruct the input and predict a pixel-by-pixel semantic segmentation. Resulting from the encoder's contractive operations and partitioning the two halves is a compressed, lower-dimensional ``bottleneck'' which forces the network to learn a compression of the overall data and learn those features most imperative to the decoder reconstruction predictions.

 The U-Net remains popular in a number of current segmentation applications due to its robustness, simplicity, and ability to more readily propagate contextual information through the entirety of the network \cite{cciccek20163d, punn2022modality}. This is accomplished through three means: ~a) an increase in the number of convolutional channels over traditional FCNNs, largely due to depth that U-Nets achieve, ~b) successive max-pooling of the data between network layers local features more-easily correlated with behavior and context at differing length scales \cite{noh2019scale}, and ~c) channel-wise concatenations of encoder feature map outputs to the decoder layers. These long-reaching concatenations, know as skip connections, decouple the encoder and decoder halves, allowing for an aggregation of multi-scale feature representation at different network stages \cite{kumar2018u, drozdzal2016importance, noh2019scale} and helping alleviate the vanishing gradient problem which plagues deeper networks \cite{ioffe2015batch}.

\paragraph{Mixed-Scale Dense Networks}
\label{sect:msdnet}

\begin{figure} [h]
\centering
    \includegraphics[width=.95\textwidth]{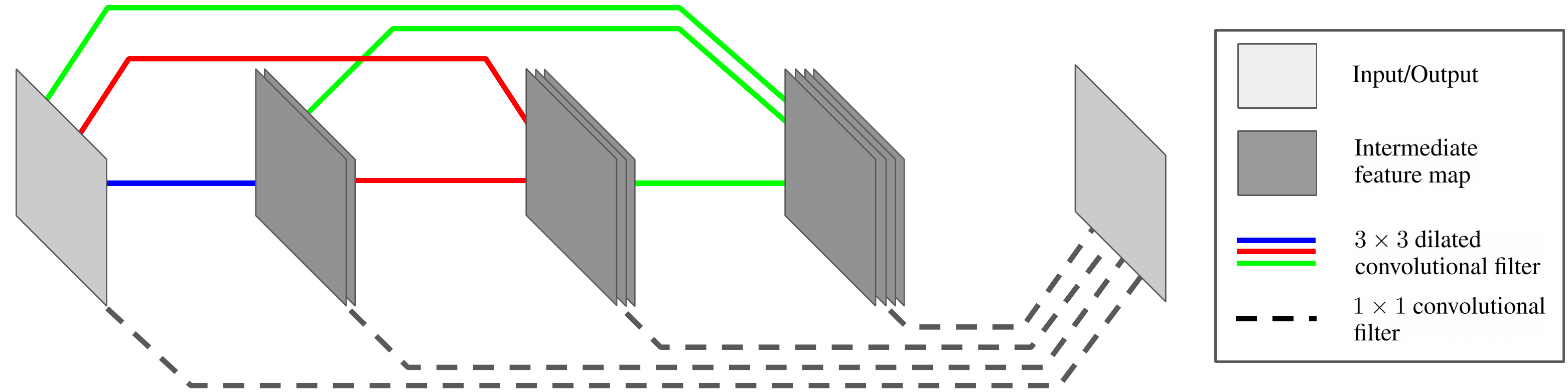}
    \caption{Schematic of a three-layer mixed-scale dense network (MSDNet). Blue, green, and red lines above represent $3\times3$ dilated convolutions between each possible pair of input and intermediate layers. Different dilations are assigned to each color. Black dotted lines below represent the $1\times 1$ convolutional operator between the output and all other layers, amounting to a linear combination with learnable weights of all input and intermediate feature maps.}
    \label{fig:msdnet}
\end{figure}

While U-net architectures remain popular, common implementations often require over several million trainable parameters which can lead to overfitting problems and harm network robustness, especially in applications where the amount of training data is low \cite{goodfellow2016deep, srivastava2014dropout}. In response, the MSDNet \cite{pelt2018mixed, pelt2018improving} architecture, depicted in Fig.~\ref{fig:msdnet}, was developed as a deep learning framework containing fewer trainable parameters (typically two to three orders of magnitude \emph{fewer}) than U-Nets. This is accomplished by densely connecting \emph{all} network layers to encourage maximum reusability of image features and by replacing the typical scaling operations found in encoder-decoder networks with dilated convolutions \cite{yu2015multi} in order to probe images at different length scales. By assigning a specific dilation to each MSDNet layer, the network can learn which dilation combinations are most effective. As a result, the number of network layers and the maximum integer dilation in which to cycle through are the most significant hyperparameters in which to toggle, drastically simplifying network design. Additionally, the dense connections among intermediate feature maps creates skip connections of \emph{all} possible lengths. Lost spatial information is more readily recovered with the inclusion of these dense skip connections, which furthermore helps alleviate the vanishing gradient problem that plagues deep networks \cite{ioffe2015batch}.


\subsection{Training Parameters}

During the model training phase, we use the multi-class cross entropy loss metric to measure how well the models classify each pixel to its respective class. The popular ADAM optimizer \cite{kingma2014adam} was chosen to minimize the loss and update the neural network weights accordingly. As for the network learning rates, all neural networks were trained for 200 epochs with an initial rate of $10^{-1}$ that was dropped by a factor of ten midway through training. For each trained model, a subset of 10\% of training data was set aside as a validation set for cross-correlation purposes and to monitor model overfitting. The network weight set correlating to the epoch with the lowest validation set loss was chosen. Lastly, each model was trained on single Nvidia RTX 3090 GPU with $24$ GB memory capacity and $936$ GB/second bandwidth using a 10 core/20 thread Intel i9-10900X CPU.

\subsection{Evaluation Metrics} \label{sect:evaluation}

To gauge model segmentation predictions in both the training and validation data sets, we use the F1 score, a popular measure of classifier performance \cite{chinchor1993muc}. It is defined as the harmonic mean of the model prediction's precision and recall, given by:

\begin{equation} \label{eq:f1}
    \text{F1} = 2*\frac{\text{precision} * \text{recall} }{\text{precision} + \text{recall}}.
\end{equation}

\begin{figure} [!htb] 
    \centering
    \includegraphics[width=.3\textwidth]{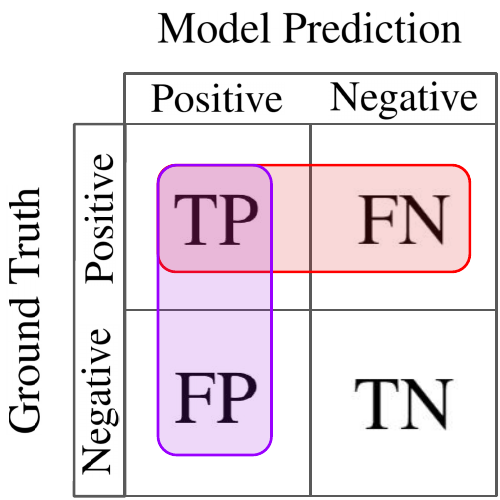}
    \caption{Confusion matrix of model predictions vs. the actual ground truth labels. Constituents of precision are highlighted vertically in purple, while constituents of recall are highlighted horizontally in red.}
    \label{fig:confusion}
\end{figure}

\noindent To further elaborate on this metric, it is useful to focus on the model predictions within each of the individual classes; we denote TP$_i$  (true positive) as the number of pixels within a class $i$ correctly identified by the model, FP$_i$ (false positive) as the number of pixels incorrectly predicted to belong in class $i$, and FN$_i$ (false negative) as the number of pixels belonging to class $i$ that are misclassified by the model. The confusion matrix diagrams these entities in Fig.~\ref{fig:confusion}. The model precision and recall within a single class $i$ is then given by:

\begin{equation} \label{eq:precision_recall} 
    \text{precision}_i = \frac{\text{TP}_i }{\text{TP}_i + \text{FP}_i}, \quad\quad\quad\quad
    \text{recall}_i = \frac{\text{TP}_i }{\text{TP}_i + \text{FN}_i}.
\end{equation}

Precision and recall are often at odds with each other, as increasing recall (the ratio of how many instances within a particular ground truth class were correctly predicted) often reduces precision (the accuracy among all model predictions made of a single class), and vice-versa. For example, an overzealous classifier may over-predict, correctly identifying most instances of certain class but erroneously producing many more false positives, leading to suitable recall but poor precision. To alleviate this, the F1 metric offers a suitable balance between the two.

To calculate the F1 score for the entire model, individual F1 scores are calculated for each individual class from their respective precision and recall metrics in Eq.~\ref{eq:precision_recall}. The full model F1 score, our target evaluation metric in Eq.~\ref{eq:f1}, then results from averaging each individual class score. To adjust for class imbalance, we compute the micro F1 score which aggregates the class scores by weighing each one differently according to relative size; i.e. the ratio of pixels belonging in each class to the total number of pixels in the dataset.

\subsection{Software Availability}
All scripts used in this project are available on GitHub (\url{https://github.com/nirajmg/FIBSEM_segmentation}). Information on how to use each file is provided in the \texttt{README} file that is part of this repository. Additionally, all neural networks were implemented using the Python-based \emph{pyMSDtorch} deep learning software library (\url{https://pymsdtorch.readthedocs.io}), On the \emph{pyMSDtorch} platform, U-Nets and MSDNets are enhanced by allowing the specification of network architecture-defining hyperparameters, such as the number of network layers, initial number of channels, convolutional channel growth rate, and custom sets of MSDNet dilations. This level of user-defined custom  which allows one to easily tune network hyperparameters to optimize performance.

\section{Overall Workflow} \label{sect:workflow}

\begin{figure} [!htb] 
    \centering
    \includegraphics[width=.99\textwidth]{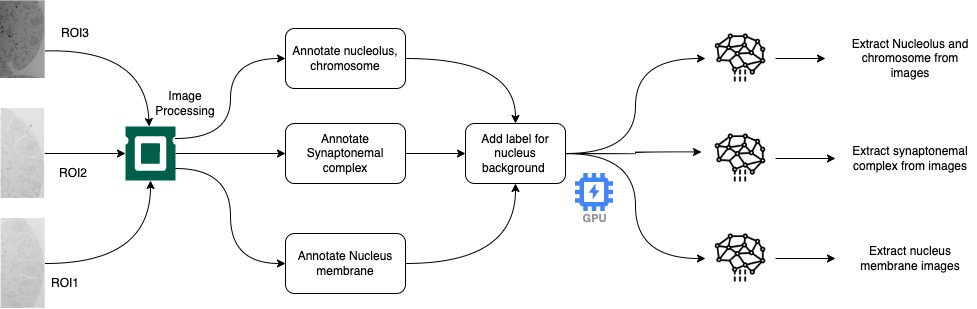}
    \caption{Complete end-to-end workflow for the segmentation of FIB-SEM data. }
    \label{fig:workflow}
\end{figure}

The ROI1, ROI2, and ROI3 nuclei tiff stacks were divided into a single nuclei jpeg file. These images were pre-processed and annotated according to Sect.~\ref{sect:preprocessing} and diagrammed in the left half of workflow chat in Fig~\ref{fig:workflow}. Only nuclei images from ROI1 are used for training the various networks. This was due to the contrast between the three stacks being remarkably similar.

The segmentation process was split into three network classifiers for extracting different groups of intracellular structures. Separating the classifiers improved segmentation accuracy and yields better results over training a single network for segmenting for all classes. Included in the training data for the first extractor model are four total annotations: one for the nucleolus, another for the chromosomes, and two background annotations, namely the interior nucleus background and exterior background located outside of the nucleus. The inclusion of two background classes greatly improved nucleolus and chromosome segmentation, as the exterior background class had a vastly different contrast and could be ignored in the gradient calculations and model parameter update steps. For the second segmentation model, the synaptonemal complex class was predicted, with prior synaptonemal complex annotations added to the aforementioned four annotations. Identifying the synaptonemal complex structures proved difficult, as their contrast was strikingly more homogeneous with the nucleus background than the other structures (blue structures in Fig.~\ref{fig:annotations}). For the third and final network, the nucleolus and nuclear membrane were the only background elements in the third batch of training data. Lastly, once all network training was complete and new segmentation predictions were inferred from the trained networks, we performed a post-processing step using skimage \cite{van2014scikit}, a Python-based open-source image library, to filter out small, superfluous objects from each individual class. More specifically, small objects of volume 1000, 2000, and 3000 pixels were filtered out for the synaponemal complex, chromosome, and nucleolus classes, respectively.

\section{Results}\label{sect:results}

We perform a parameter sweep on the U-Net architecture-governing hyperparameters to find the best performing neural network classifiers for segmenting the target cell structures consisting of the nucleolus, chromosomes, synaptonemal complex, and the cellular membrane. The hyperparameters considered in this analysis were the depth of the U-Net, number of initial base channels, interlayer growth rate of convolutional channels, and batch size. Results of the U-Net model sweep for the three separate classifiers are shown in Table \ref{tab:table1}, where for each column, the same hyperparameters are used for the three networks. Alternatively. the parameter sweep results for the single classifier accommodating all classes are shown in Table \ref{tab:table2}. For both, the best performing networks with respect to the micro F1 evaluation score, referenced and described in Sect.~\ref{sect:evaluation}, are highlighted in gray. 

    \begin{table} [!htb]  \small
        \parbox{.47\linewidth}{
            \centering
            \begin{tabular}{c|cccc>{\columncolor[gray]{0.8}}ccccc}
            \hline \\[-1em]
            Model                               & 0       & 1       & 2       & 3       & 4      \\
            \hline \hline\\[-1em]
            Depth                               & 4       & 4       & 4       & 5      & 5      \\
            \hline\\[-1em]
            \begin{tabular}{@{}c@{}} Base  channels   \end{tabular}    & 32      & 64      & 64      & 32     & 64      \\
            \hline\\[-1em]
            \begin{tabular}{@{}c@{}}Growth  rate \end{tabular}     & 2     & 2       & 1.5      & 2       & 2       \\
            \hline \hline\\[-1em]
            \begin{tabular}{@{}c@{}}Parameter \\ count ($10^6$)\end{tabular}     
                                                & 2.14   & 8.56   & 2.57  & 8.63   & 34.51  \\
            \hline \hline\\[-1em]
            \begin{tabular}{@{}c@{}}Training \\ loss ($10^{-2}$)\end{tabular}                    
                                                & 2.86 & 2.94 & 2.97 & 2.73 & 2.78 \\
            \hline\\[-1em]
            \begin{tabular}{@{}c@{}}Validation \\ loss ($10^{-2}$)\end{tabular}               
                                                & 3.19 & 3.3 & 3.42 & 3.40 & 3.45 \\
            \hline\\[-1em]
            \begin{tabular}{@{}c@{}}Micro F1 \\ score \end{tabular}  
                                                & .902  & .902  & .896  & .898  & .904  \\
            \hline
            \end{tabular}
            
            \vspace{10pt}
            \caption{Summary of hyperparameter sweep results for U-Nets trained on data with nucleolus and chromosome \emph{only}. The best performing network according to the F1 evaluation metric is highlighted in gray. \label{tab:table1}}
        }
    \hfill
        \parbox{.47\linewidth}{
            \begin{tabular}{c|c>{\columncolor[gray]{0.8}}cccccccc}
            \hline\\[-1em]
            Model                               & 0       & 1       & 2       & 3       & 4      \\
            \hline \hline\\[-1em]
            Depth                               & 4       & 5       & 4       & 4      & 5      \\
            \hline\\[-1em]
            \begin{tabular}{@{}c@{}} Base  channels   \end{tabular}    & 32      & 32      & 64      & 64      & 64      \\
            \hline\\[-1em]
            \begin{tabular}{@{}c@{}}Growth  rate \end{tabular}     & 2     & 2       & 1.5      & 2       & 2       \\
            \hline \hline\\[-1em]
            \begin{tabular}{@{}c@{}}Parameter \\ count ($10^6$)\end{tabular}     
                                            & 2.14   & 8.63   & 2.57  & 8.56   & 34.51  \\
        
            \hline \hline\\[-1em]
            \begin{tabular}{@{}c@{}}Training \\ loss ($10^{-2}$)\end{tabular}                    
                                                & 3.617 & 3.82 & 3.89 & 3.55 & 3.41 \\
            \hline\\[-1em]
            \begin{tabular}{@{}c@{}}Validation \\ loss ($10^{-2}$)\end{tabular}               
                                                & 3.99 & 3.88 & 4.11 & 3.97 & 3.87 \\
            \hline\\[-1em]
            \begin{tabular}{@{}c@{}}Micro F1 \\ score \end{tabular}  
                                                & .766  & .781  & .7445 & .769  & 0.781  \\
            \hline
            \end{tabular}
            
            \vspace{10pt}
            \caption{Summary of hyperparameter sweep results for the U-Nets trained on \emph{all} labeled classes, including synaptonemal complex. The best performing network is once again highlighted in gray. \label{tab:table2}}
        }
    \end{table}

Upon inspection, the variation in depth or base channels has little effect on training and validation losses in the multi-network classifiers in Table \ref{tab:table1}. According to the F1 scores, all U-Net configurations perform within 1\% of each other, which is certainly a testament to the robustness of U-Net segmentation schemes. With a training F1 score of 0.904, the U-Net Model 4 with 64 base channels and a depth of 5 produce the best results. In contrast, for the single-network classifiers in Table \ref{tab:table2}, variation in the number of base channels and depth has a more pronounced impact on the variability of segmentation results. F1 scores are within 5\% of each other. Here, with a score of 0.781, the Model 1 with 32 base channels and depth 5 is the best performing configuration. No sweep was performed for the MSDNet classifier, as memory constraints limited this study to MSDNet layer depth of 40 and a maximum dilation setting of 15, considerably lower than the 100- and 200-layer networks implemented in the original paper and subsequent applications \cite{pelt2018improving, pelt2018mixed, zeegers2020task}. MSDNet performance was satisfactory with a validation F1 score of 0.8773, despite this constraint, However, this mark falls generally below the single network U-Net classifiers in Fig.~\ref{tab:table1}.  

\begin{figure}[!htb]
    \minipage{0.33\textwidth}
        \centering
        \includegraphics[width=5cm,height=5cm]{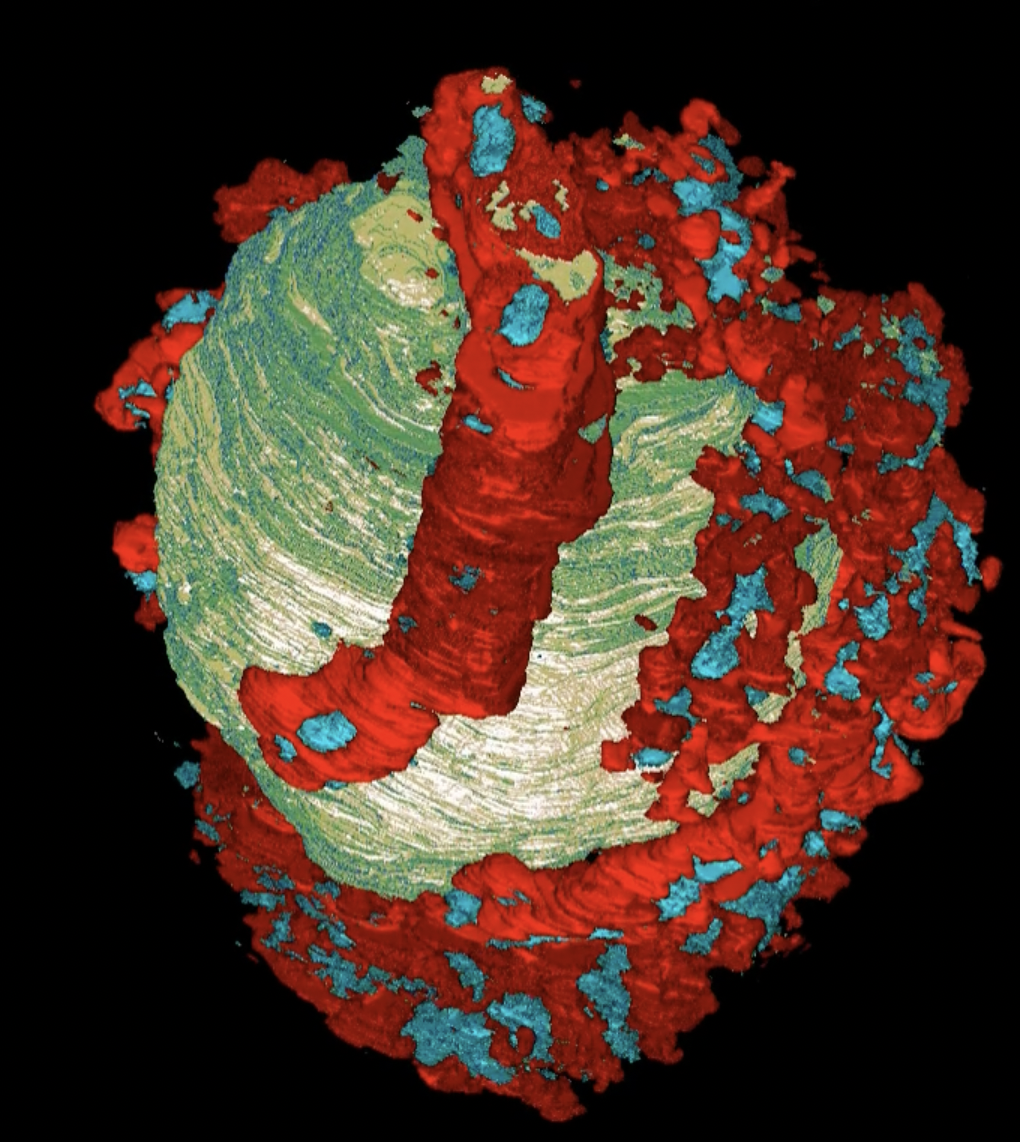}
      
    \endminipage\hfill
    \minipage{0.33\textwidth}
        \centering
        \includegraphics[width=5cm,height=5cm]{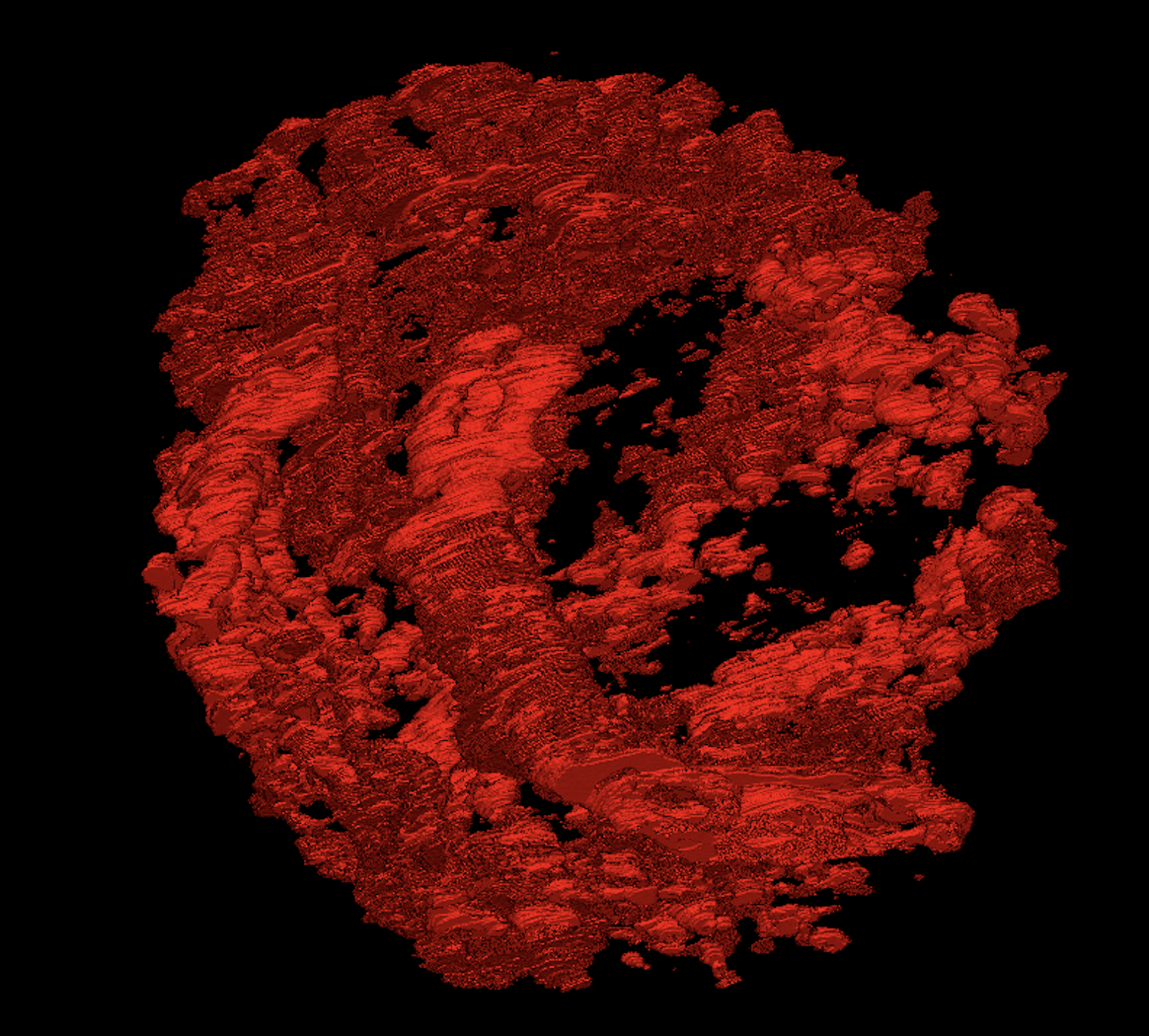}
      
    \endminipage\hfill
    \minipage{0.33\textwidth}%
        \centering
        \includegraphics[width=5cm,height=5cm]{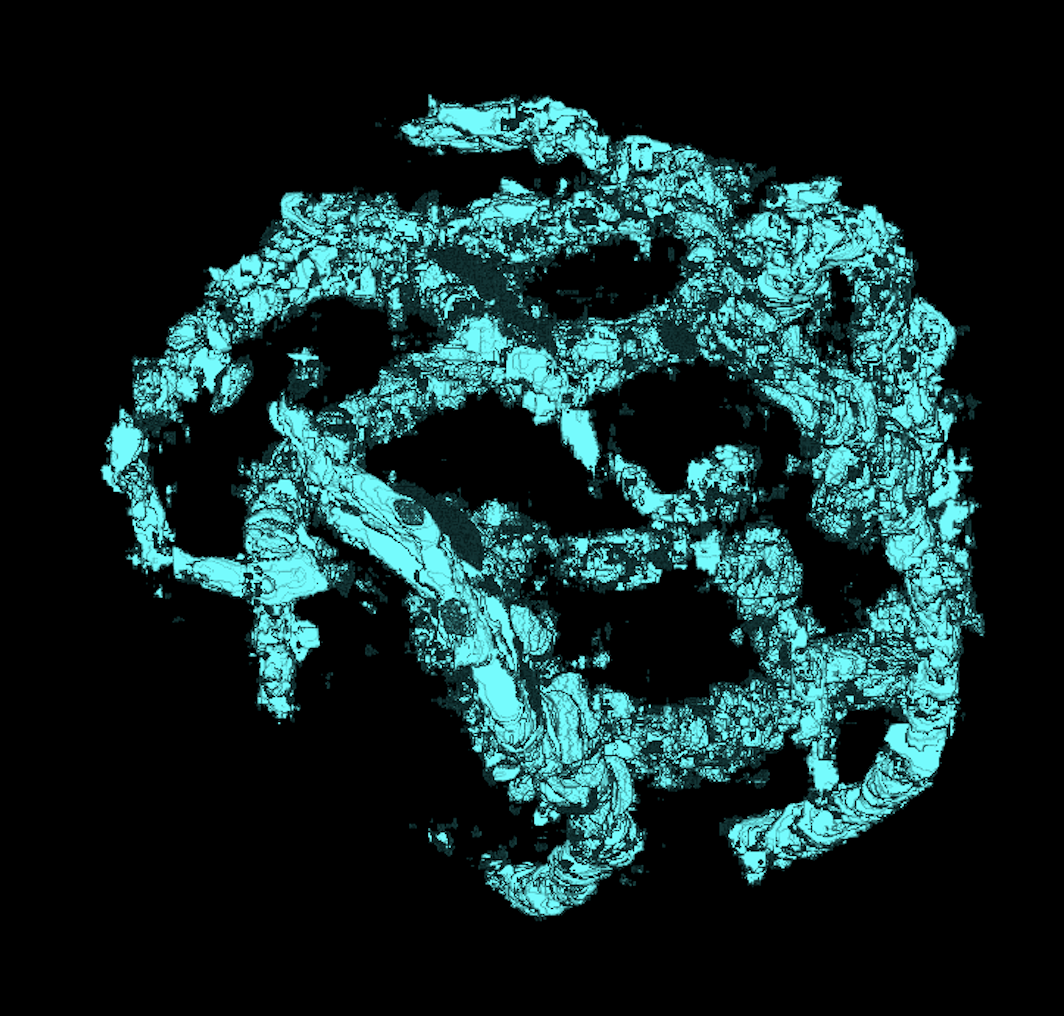}
     
    \endminipage
     \caption{ROI1 stack segmentation results.The first image displays the 3d orientation of nucleolus, chromosome and synaptonemal complex in a nucleus.  The synaptonemal complex is seen in the Third image, whereas the second image shows a three-dimensional representation of a chromosome.}\label{fig:ROI1}
\end{figure}
\begin{figure}[!htb]
\minipage{0.33\textwidth}
  \centering
    \includegraphics[width=5cm,height=5cm]{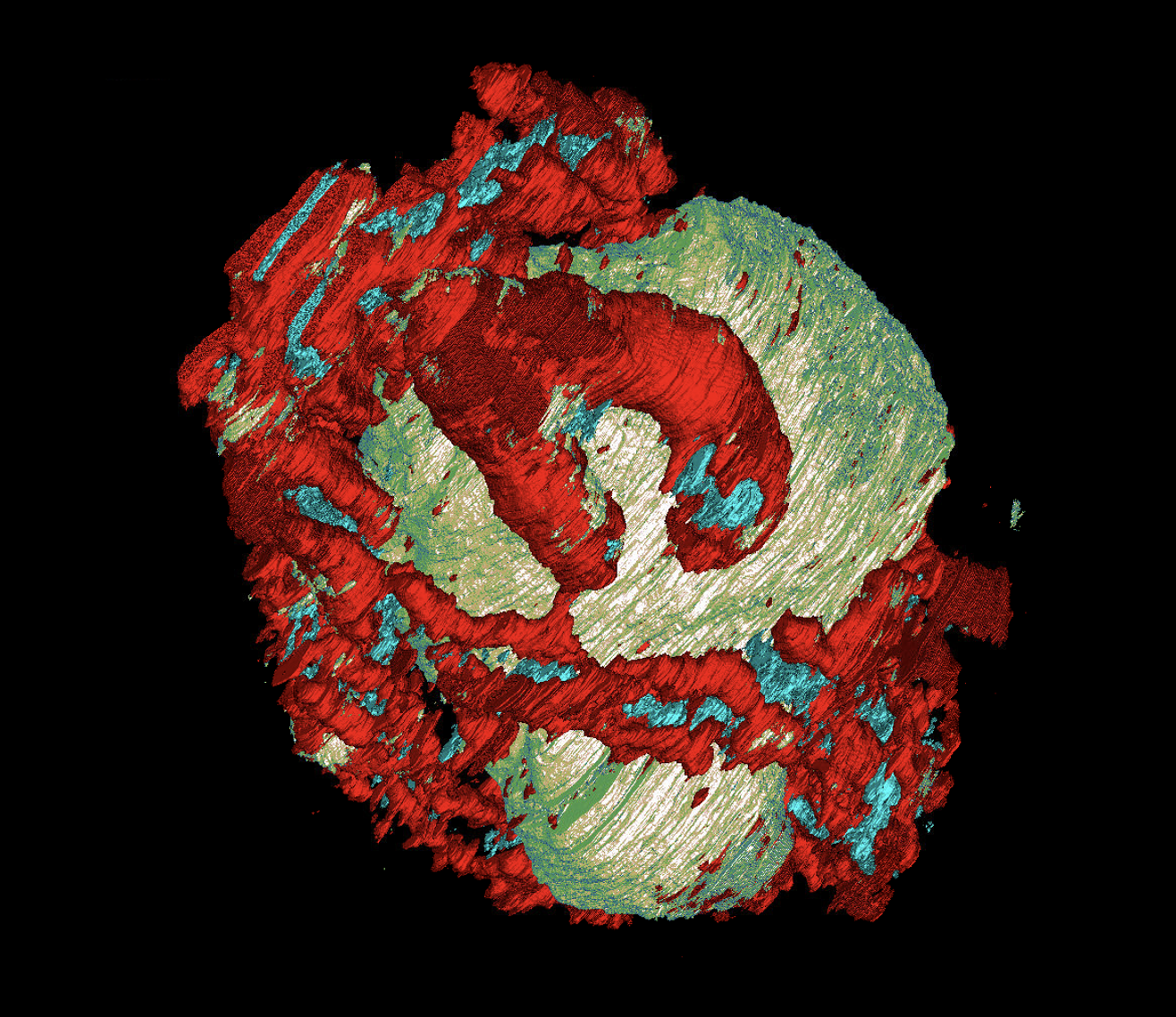}
  
\endminipage\hfill
\minipage{0.33\textwidth}
    \centering
    \includegraphics[width=5cm,height=5cm]{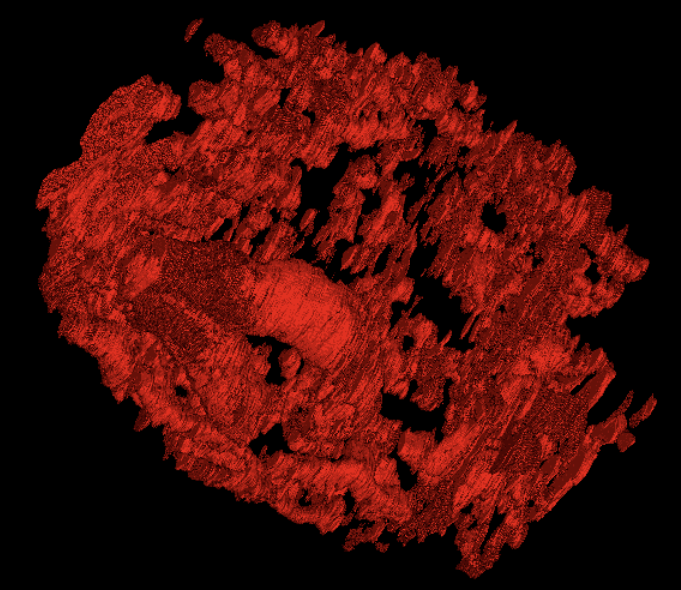}
  
\endminipage\hfill
\minipage{0.33\textwidth}%
    \centering
    \includegraphics[width=5cm,height=5cm]{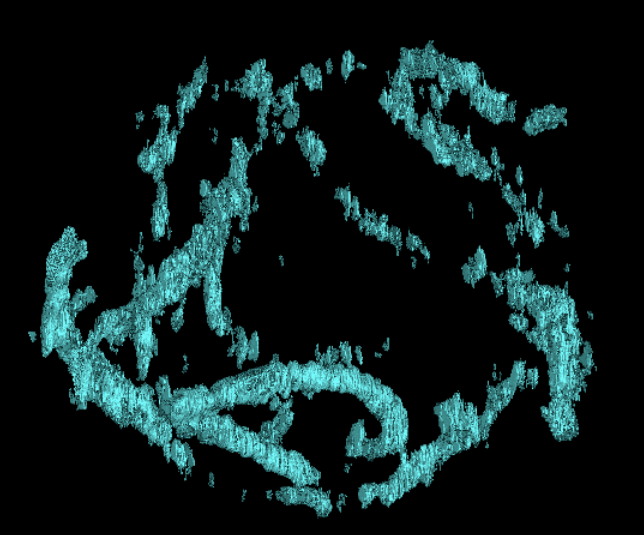}
 
\endminipage
 \caption{ROI2 stack segmentation results.The first image displays the 3d orientation of nucleolus, chromosome and synaptonemal complex in a nucleus.  The synaptonemal complex is seen in the Third image, whereas the second image shows a three-dimensional representation of a chromosome.}\label{fig:ROI2}

\end{figure}

 We focus primarily on results from the multi-network classifiers described in Sect.~\ref{fig:workflow}, particularly the Model 4 configuration in Table \ref{tab:table1}. All of the networks are primarily trained on ROI1 images using random samples from various nuclei in ROI1. This is done to guarantee the full range of image contrast is represented during the network phase. Figure \ref{fig:ROI1} depicts the segmentation results for ROI1 from the Model 4 U-Net configuration. The left-most graphic depicts a 3D volumetric image the first two network classifier results, one trained to segment the cell chromosomes (in red) and nucleolus, and another trained to segment the synaptonemal complex (in blue). The middle subimage displays only the chromosome segmentations, while the right-most subimage shows the synaptonemal complexes segmentations, independent from their chromosome encasing. The synaptonemal complex segmentation remains particularly impressive, as their contrast is nearly homogeneous with the nuclear background. This result highlights the strength and generalizability of neural network-based segmentation. While the ROI1 segmentation in Fig.~\ref{fig:ROI1} results from a U-Net trained on small subset of 160 images from ROI1 itself, no images from ROI2 were present in the training data or network learning process. The networks were completely blind to ROI2 data. The resulting segmentation of ROI2 nucleolus, chromosomes, and synaptonemal complexes is shown in Figure \ref{fig:ROI2}.
 

\begin{figure}[!htb]
\begin{minipage}{.45\textwidth}
            \centering
          \includegraphics[width=7.2cm,height=7.2cm]{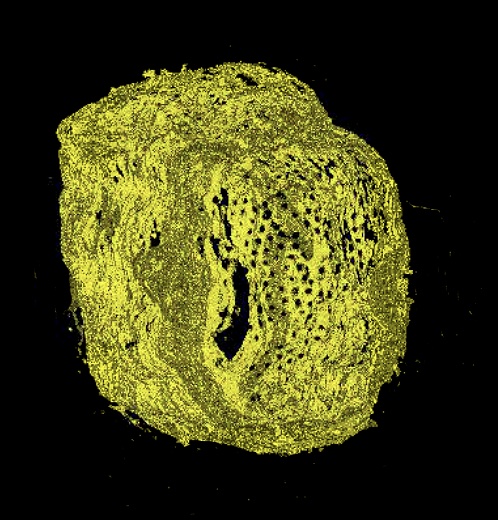}
         \caption{Segmentation results of nuclear membrane from ROI1.}
         \label{fig:membrane}
     \end{minipage} \hfill
\begin{minipage}{.45\textwidth}
        \centering
    
          \includegraphics[width=7.2cm,height=7.2cm]{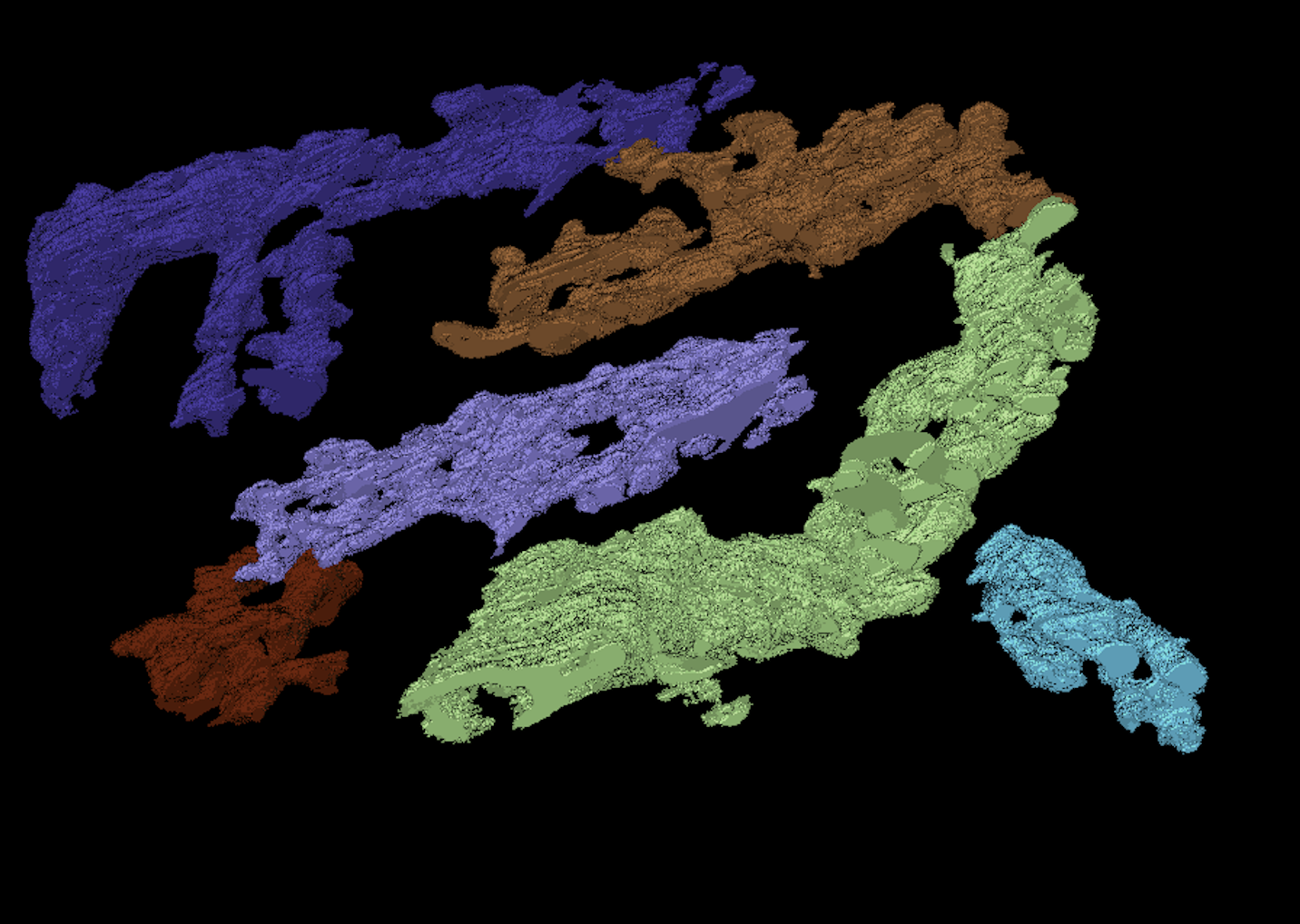}
         \caption{Extraction of individual chromosome from ROI1 using the 3d connected components.}
         \label{fig:chromosomes}
     \end{minipage}
\end{figure}

 The third network classifier identifies the nucleus membrane, pictured in Fig.~\ref{fig:membrane}. Gaps in the membrane walls appear, those these gaps are expected as they are artefacts resulting from imaging post-processing steps. Additionally, Fig.~\ref{fig:chromosomes} displays individually connected chromosomes that are labeled using 3D connected components. By setting the connectivity to 26, which constructs the decision tree for labeling decisions, and delta to 10, which dictates that any nearby voxel value less than 10 is treated as the same component, the segmented chromosome's twelve biggest components can be identified using the cc3d library \cite{rosenfeld1966sequential, sutheebanjard2012decision, silversmith2021cc3d}.
 

%

\section{Discussion and Future Work} \label{sect:discussion}

In this study, we apply machine learning-based automated  on large swaths of FIB-SEM nucleus data of C. elegans gonad eggs, providing pixel-by-pixel segmentation of all sub-nuclear cellular structures, including the nucleolus, chromosomes, synaptonemal complexes enveloped in chromosome pairs, and the nucleus membrane. We achieve an impressive micro F1 classification scoring metric to 0.904 and provide valuable insight into the morphological analysis and contextual insights with regard to cellular processes. We employ two deep learning architectures: U-Net and mixed-scale dense networks.  Particularly impressive are the networks' ability to learn the synaptonemal complex pixels; these structures had strikingly similar contrast to that of the nucleus background. We hypothesize that the networks learned not only from the complexes' visual patterns, but also from the added context that the complexes were almost entirely encased in dual chromosome strands. In total, we analyzed a sizable amount of isotropic volumetric data; 6000 cross sectional slices of size $2500\times5000$ pixels spread across three regions of interest (ROIs), though the machine learning classifiers were trained on smaller cross sectional image slices encapsulating single nuclei at varying depths, each sized at $1700\times1700$ pixels. Impressively, this data was resolved to voxels of size $4 \times 4 \times 4$ \si{\nano\meter}$^3$. In order to generalize the training and segmenting of such large images onto smaller platforms, our approaches offer batch segmentation of images to reduce memory costs.


For future work, we envision deep learning models incorporating full 3D volumes of images. Though memory intensive, 3D deep learning methods may be able to better contextualize local information given the extra neighboring information in higher dimensionality. A challenge here remains, namely the difficulty in curating 3D masks and labels for a large and representative training set of images. However, this difficulty may be alleviated with the use of mixed-scale dense networks (MSDNets, detailed in Sect.~\ref{sect:msdnet}). Though MSDNets performed sub-optimally compared to U-Nets in this current study, their densely-connected architecture and maximum reusability of data was specifically designed to perform better on smaller training data sets and sparse labeling \cite{pelt2018mixed, pelt2018improving}, and their ability to accommodate 3D volumes of images with little network architecture is advantageous,  Alternatively, patch-based (or tile-based) data augmentation schemes in which smaller, overlapping subsets of images are drawn from the original data \cite{innamorati2019learning, cui2019deep} may be generalized to 3D volumes of data. allowing 3D U-Nets and MSDNets to learn from significantly smaller sets of labels and training data, evidenced by similar overlap averaging techniques \cite{pielawski2020introducing}.

The contrast disparity between different ROI stacks remains a difficulty, as the network performs sub-optimally on ROI3. This necessitates either retraining the network on a sample of photos or pre-processing of images to match the contrast with the training images. In this instance, one of the methods utilized to match the ROI3 contrast with the ROI1 and ROI2 was histogram matching. To generalize our workflow to cryo-electron microscopy (cryo-EM) data, we intend to use similar techniques. Cryo-EM, a new biophysical method for determining the structure of protein complexes, is becoming more and more popular. Recognizing the sophisticated molecular features on medium-resolution cryo-EM density maps is difficult, though experimental cryo EM structures could be segmented more efficiently by applying deep learning to them.

\section{Funding}

This work was partially funded by NIH award number 5R00GM132544-04 and University of Colorado, Boulder start-up funds belonging to V. K. Further support originates from the National Institute of General Medical Sciences of the National Institutes of Health (NIH) under Award 5R21GM129649-02 and from the Laboratory Directed Research and Development Program of Lawrence Berkeley National Laboratory under U.S. Department of Energy Contract number DE-AC02-05CH11231. 




\section{Author Contributions}

VK and PHZ supervised the project; NG, EJR, SP, CSX, HFH, and VK wrote the manuscript (original draft); VK, PHZ, NG, and EJR reviewed and edited the manuscript;  CSX, SP, and HFH collected FIB-SEM data; FW, DG, and VK
prepared C elegans samples. FW and AD provided C. elegans strains and
advised on the study; NG and EJR performed data analysis and machine learning architecture design and training; EJR and PHZ designed pyMSDtorch machine learning software suite; NG and VK uploaded data and provided workflow to the GitHub repository; PHZ proposed the machine learning solutions;  VK proposed the biological questions and conceived the study.


\bibliographystyle{unsrt}  

\bibliography{references}  

\end{document}